\documentclass{aa}
\usepackage{graphicx}
\usepackage{txfonts}
\usepackage{amsmath}
\usepackage{natbib}
\bibpunct{(}{)}{;}{a}{}{,} 
\usepackage{graphicx}
\usepackage{longtable}
\usepackage{supertabular}
\usepackage{hyperref}
\begin{document}

\title{The inner two degrees of the Milky Way }
\subtitle{Evidence of a chemical difference between the  Galactic Center and the surrounding inner bulge stellar populations}
\author{M. Schultheis\inst{1}
\and    R.~M.~Rich \inst{2}  
\and L. Origlia \inst{3}
\and N. Ryde \inst{4}
\and G.~Nandakumar \inst{5,6}
\and B. Thorsbro \inst{4}
\and N. Neumayer \inst{7}
}

 \institute{Universit\'e C\^ote d'Azur, Observatoire de la C\^ote d'Azur, CNRS, Laboratoire Lagrange, Blvd de l'Observatoire, F-06304 Nice, France
 e-mail: mathias.schultheis@oca.eu
\and
Department of Physics and Astronomy, UCLA, 430 Portola Plaza, Box 951547, Los Angeles, CA 90095-1547, USA; rmr@astro.ucla.edu
\and
INAF-Osservatorio di Astrofisica e Scienza dello Spazio, Via Gobetti 93/3, I-40129 Bologna, Italy
\and
Lund Observatory, Department of Astronomy and Theoretical Physics, Lund University, Box 43, SE-221 00 Lund, Sweden
\and
Research School of Astronomy \& Astrophysics, Australian National University, ACT 2611, Australia
\and
ARC Centre of Excellence for All Sky Astrophysics in Three Dimensions (ASTRO-3D)
\and
Max-Planck-Institut für Astronomie, Königstuhl 17, 69117 Heidelberg, Germany
}

\abstract
{ Although there have been numerous studies of chemical abundances in the Galactic bulge, the central two degrees have been relatively unexplored due to the heavy and variable interstellar extinction, extreme stellar crowding, and the presence of complex foreground disk stellar populations. }
{In this paper we discuss the  metallicity distribution function, vertical and radial gradients and  chemical abundances of $\alpha$-elements in the inner two degrees of the Milky Way, as obtained by recent IR spectroscopic surveys.}
{ We use a  compilation of recent measurements of metallicities and $\alpha$-element abundances derived from medium-high resolution spectroscopy. We compare these metallicities with low-resolution studies.} 
{ Defining "metal-rich" as stars with $ \rm [Fe/H]>0$, 
and "metal-poor" as stars with $\rm [Fe/H]<0$, we find compelling evidence for a higher fraction ($\sim 80\%$) of metal-rich stars in the Galactic Center (GC) compared to the values (50-60\%) 
measured in the low latitude fields within the innermost 600\,pc. The high fraction of metal-rich stars in the GC region  implies a very high mean metallicity of +0.2\,dex,  while in the inner 600 pc of the bulge the mean metallicity is rather homogenous around the solar value. A vertical metallicity gradient of -0.27\,dex/kpc in the inner 600\,pc is only measured if the GC is included, otherwise the distribution is about flat and  consistent with  no vertical gradient.}
{ In addition to its high stellar density, the Galactic center/nuclear star cluster is also extreme in hosting high stellar abundances, compared to the 
surrounding inner bulge stellar populations; this has implications for formation scenarios and strengthens the case for the NSC being a distinct stellar system.}



\titlerunning{The central inner two degrees of the Milky Way}
\authorrunning{Schultheis et al.}
\maketitle

\section{Introduction}
\begin{figure*}
    \centering
    \includegraphics[width=1.0\textwidth]{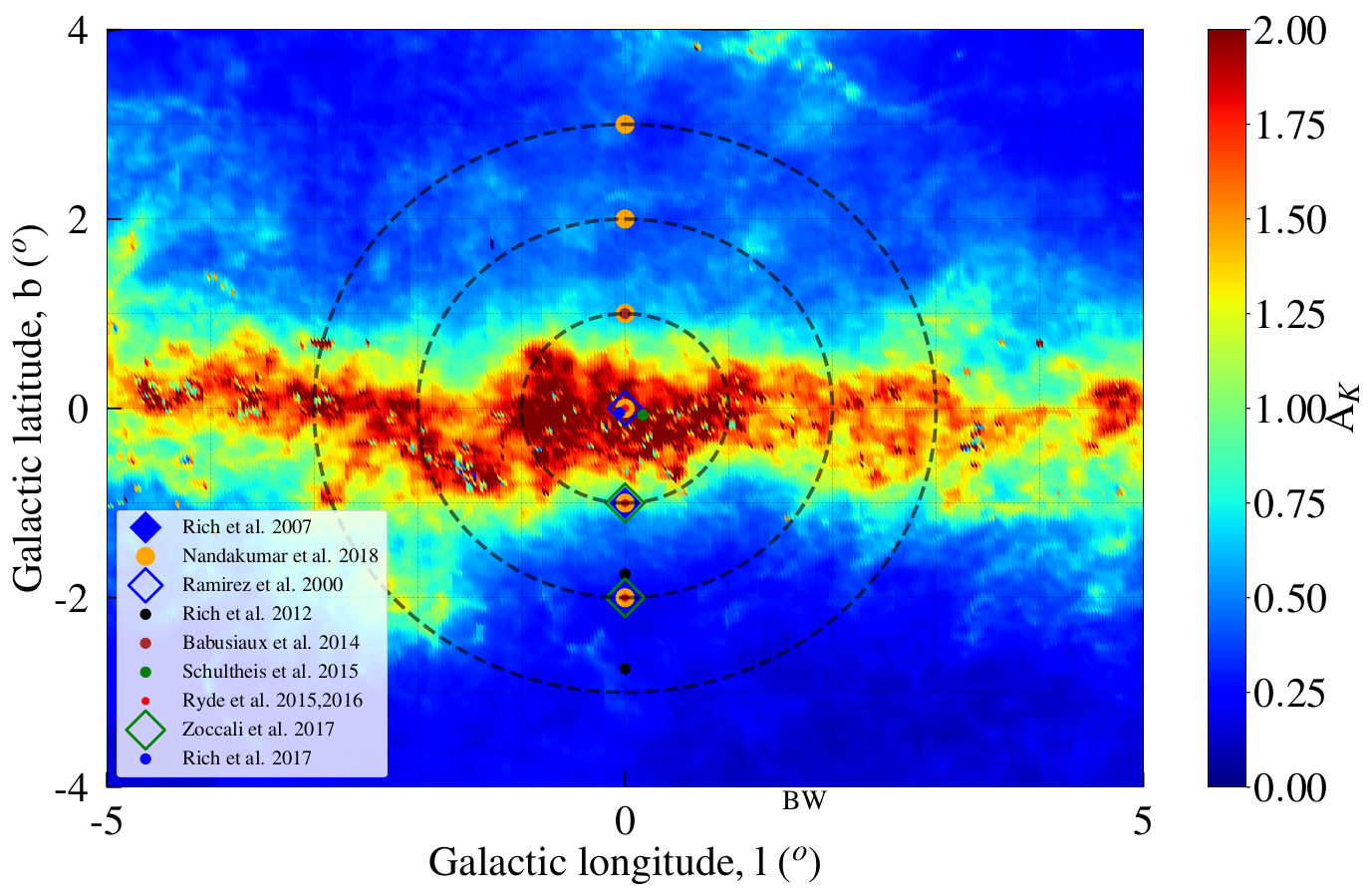}
    \caption{Location of target fields from the spectroscopic studies in the inner bulge, superimposed on the interstellar extinction map of \protect \citet{gonzalez2012}. The dashed lines  denote an angular radius of  1,2 and 3 degrees from the Galactic Center, respectively.}
    \label{Fig1}
\end{figure*}

The era of large spectroscopic surveys has strengthened the case that the central regions of the galaxy are dominated by an old metal-rich bulge/bar stellar population that is distinct in formation history from the thin disk (see e.g. the reviews of \citealt{rich:13}, \citealt{Origlia2014}, and \citealt{Barbuy2018}). 
The vast majority of these studies have been done at optical wavelengths $\rm (<1\,\mu m)$
and in regions of relatively low interstellar extinction at $b<-2^o$, such
as the well-known Baade window, although recent efforts have probed the bulge closer to the plane (\citealt{zoccali2017}, \citealt{trapp2018}). In these regions, due to the large and high variable interstellar extinction (see e.g. \citealt{gonzalez2012}, \citealt{schultheis2014}, \citealt{Nogueras-Lara2018a}), observations are limited to the near-infrared and longer wavelengths.  With the recent development of high resolution infrared spectrographs, it is now possible to get detailed chemical abundances of giant stars in the inner Galactic bulge, although cool giants (with substantial molecular lines) remain a challenge for abundance analysis.   We define here the inner Galactic bulge  with $\rm |l| < 10\degr, |b| \leq 3\degr$ assuming a distance to the Galactic Center of 8.2\,kpc (\citealt{Karim2017}).\footnote{ 1 degree corresponds to 145\,pc.} Early high resolution infrared spectroscopy commenced with nirspec on Keck II e.g. \citet{rich:07,rich:12}, and then on larger samples with the APOGEE survey (\citealt{Majewski2017}).  As was the case in the earlier studies with nirspec, APOGEE works in the $H$-band where extinction can still be significant ($\rm A_{V}/A_{H} \sim 6$), especially in the inner degree where it reaches extreme values (see e.g. \citealt{schultheis2009}, \citealt{fritz2011}, \citealt{Nogueras-Lara2018a}). $K$-band spectroscopy ($\rm A_{V}/A_{K} \sim 10$) of  M giants is thus the most efficient way to measure detailed chemical signatures in the innermost regions (see e.g. \citealt{cunha2007}, \citealt{ryde_schultheis:15}, \citealt{Nandakumar2018}). Figure~\ref{Fig1} shows a view of the different studies obtained in the inner Galactic bulge superimposed to the extinction map of \citet{gonzalez2012}.

In the central 5\,pc, the stellar population is dominated by the nuclear star cluster, which hosts a wide range of stellar ages and metallicities, even including young massive stars \citep{Genzel2010}. 
NSCs are in general the most dense stellar systems in the universe, with masses of  $\rm ~ 10^{7}~M_{\odot}$  and  typical sizes of $\rm \sim 5\,pc$ (see \citealt{Neumayer2017}).
The NSC of the Milky Way has an estimated half-light radius of $\rm 4.2 \pm 0.4 pc$ \citep{Schoedel2014}  and a mass of $\rm 2\times10^{7}~M_{\odot}$  \citep{fritz16, Feldmeier2014}  and it also  hosts the supermassive black hole, $\rm Sgr A^{*}$.   Its formation mechanism is still under debate. 
Two main scenarios have been proposed: (i) Stars have formed `in situ' in the center of the Milky Way  (e.g. \citealt{Seth2008}, \citealt{Milosavljevic2004}, \citealt{Pflamm-Altenburg2009}) (ii) Star clusters have migrated from a larger distance (e.g. \citealt{Tremaine1975},  \citealt{Capuzzo-Dolcetta1993}, \citealt{Antonini2013}, \citealt{Gnedin2014}). The metallicity distribution function, as well as detailed chemical abundances, are important constraints on the formation history of the NSC and inner bulge close to the Galactic plane.
In the NSC, the cool stars amenable to abundance determination can run the gamut from red supergiants several Myr old, to red giants t$\ge$10 Gyr old.
\citet{carr2000} and \mbox{\citet{Ramirez2000}} carried out detailed abundance analysis using high resolution IR spectra ($R\sim$40,000) and estimated a mean metallicity of $+$0.12$\pm$0.22\,dex for 10 cool luminous supergiant stars in the Galactic Center (GC). The same stars were re-measured  by \cite{cunha2007} with a slightly higher resolution ($R\sim$50,000) and they estimated a similar mean metallicity. However, red supergiant stars have complex stellar atmospheres that are challenging to model and the analysis of such spectra is therefore nontrivial (see e.g. \citealt{cunha:2006}). \citet{Ramirez2000} determined, for the first time,
metallicities of M giant stars in bulge fields along the minor axis  using
low-resolution ($R\sim1300-4800$) spectra. They determined metallicities based on the equivalent widths of three strong features in their K-band  spectra, namely NaI, CaI, and  the first overtone band  of CO.
They did not find any evidence for a metallicity gradient along the minor or major axes in the inner bulge ($R_{\rm GC}\lesssim 560\,$pc).

\citet{rich:07} did one of the first detailed abundance determination from high resolution IR spectroscopy with nirspec \citep{nirspec_mclean} at KeckII, by analysing  17 M giants located at ($l$,$b$)=($0^\circ,-1^\circ$).  They found a mean iron abundance of [Fe/H]=$-0.22$ with a $\rm 1\,\sigma$ dispersion of 0.14\,dex. Later on, \citet{rich:12}  carried out a consistent analysis of 30 M giants at ($l$,$b$)=($0,-1\fdg75$) and ($1^\circ,-2\fdg75$) that were observed with the same instrument and estimated mean iron abundances of $-0.16\pm0.12$\,dex and $-0.21\pm0.08$\,dex, respectively. These authors also combined their analysis of 14 M giants in Baade's Window (l=$1\fdg02$, $b$=$-3\fdg93$) using the same instrument and \citet{rich:05} found no gradient in abundance or $\rm [\alpha/Fe]$  in the innermost 150\,pc to 600\,pc region.

\cite{babusiaux:14} determined metallicities for $\sim$100 red clump (RC) stars at ($l$,$b$) = ($0^\circ$,+$1^\circ$) using low resolution optical spectra ($R\sim$6,500) and found a mean metallicity very similar to that of \cite{ramirez:00} and \cite{rich:07} at ($l$,$b$) = ($0^\circ, -1^\circ$), 
suggesting symmetry between Northern and Southern inner bulge fields.

\citet{schultheis:15} analysed APOGEE spectra of cool M giants in the inner degree, finding a significant presence of a metal-poor population enhanced in $\alpha$-elements. 
However,  interstellar extinction limited APOGEE observations near the Galactic Center, and only some red supergiants and luminous AGB stars (see \citealt{schultheis:15}) could be measured. 

\citet{Do:2015} derived  metallicities for about 80 M giants in the NSC observed at low resolution ($R\simeq 5,000$)  with the Gemini-North's Near-Infrared Integral Field Spectrometer (NIFS).  Their Bayesian analysis reported a wide range in derived metallicities, extending to +0.96 dex and with 6\% of their sample having $\rm  [Fe/H]<-0.5$.  However, their abundances can be considered to be only estimates, as they also reported $\rm log g >3$ for most of their sample.

\citet{Feldmeier-Krause2017} used a method similar to \citet{Do:2015}, reporting approximate metallicities for 700 M giants in the vicintiy of the NSC, using VLT-KMOS spectroscopy at $R=4000$.  They also reported a suprasolar metallicity and a metal poor fraction similar to that of \citet{Do:2015}. 

\citet{Nogueras-Lara2018b} used a purely photometric result and fitted luminosity functions to their deep photometric data in the inner bulge with $\rm \alpha-enhanced$ BASTI isochrones and found that an old stellar population with Z=0.04 giving a [Fe/H]=0.18\,dex provides the best fit.

One of the first detailed abundance analyses in the GC  using  $K$-band high-resolution CRIRES at the ESO/VLT spectra were obtained by \citet{ryde_schultheis:15}. They analyzed spectra of 9 field giants in the vicinity of the NSC, finding a narrow metallicity distribution with $\rm [Fe/H]= +0.11 \pm 0.15\,dex$, in good agreement with \citet{cunha2007}. Their $\alpha$-element abundances are found to be low, following the trends from studies in the outer bulge, and resembling a bar-like population. 
A refined analysis 
 (\citealt{Nandakumar2018}) gives a slighter higher mean metallicity of $\rm [Fe/H]= +0.3 \pm 0.10\,dex$ and confirms the very narrow distribution.

\citet{ryde:16} presented an abundance study of 28 M giants in fields located within a few degrees south of the Galactic Center using high resolution ($R\sim$50,000) spectra. They found a wide range of metallicities that narrows towards the center and confirmed the \citet{rich:12} study that found alpha enhancement throughout the inner bulge. This would be consistent with a homogeneous enrichment history in the inner bulge.

Very recently, new and more massive samples of chemical abundances from the APOGEE and GIBS surveys, as well as from high resolution spectroscopy at Keck and VLT became 
available. In Sect.~\ref{samples} we will describe  this new dataset of chemical abundances and in the Sects. 3, 4, and 5 we will used this dataset to constrain the chemical enrichment of the inner bulge and Galactic Center region. 

\section{Recent spectroscopy in the inner bulge and in the Galactic Center region}
\label{samples}

\citet{rich:17} measured metallicities for 17 M giants  in the NSC. Membership of those stars is based on detailed extinction measurements as well as on kinematics, i.e. proper motions, radial velocities, and orbit calculations. 
Below, we will refer to this sample as R17.

\citet{Feldmeier-Krause2017}  used full spectral fitting to derive effective temperatures and metallicities for more than 700 M giant stars.
They found a significant fraction of metal-poor stars and  found also a significant amount of super metal-rich stars with $\rm [Fe/H] > 0.5$, somewhat ruling out the scenario that the NSC could be entirely formed from infalling of globular clusters. However, as they mentioned, these extremely metal-rich stars should be regarded with  caution. We will refer to this sample as FK17.

In the APOGEE DR12 bulge sample \cite{garciaperez:18} identified more than two components in the metallicity distribution by arranging the red giant stars according to their projected Galactocentric distance and distance from the Galactic mid-plane. \cite{fragkoudi:18}, by using the  metallicities derived from the  APOGEE DR13 compared the observed metallicity Distribution Function (MDF) with that obtained from N-body simulations of a composite (thin+thick) stellar disc. 

\citet{Zasowski2019} presented chemical abundances of M giant stars from the latest APOGEE DR14,  in the inner three degrees. We selected stars along the minor axis ($\rm l=\pm 0.2\degr$) and at galactic latitudes ($\pm 1\degr$, $\pm 2\degr$, and $\pm 3\degr$). Henceforth, we will refer to this sample as Za19.

\citet{zoccali2017}  derived metallicities for 432 stars   at $b$ = -2\degr ~and $b$ = -1\degr from the GIRAFFE Inner Bulge Survey (GIBS), using the optical Ca Infrared triplet method. They  found evidence for a  bimodal MDF. Using a similar definition of "metal rich" being suprasolar, and "metal poor" being subsolar, they argue that the "metal rich" population is flattened and concentrated toward the plane, while the "metal poor" component is spheroidal.  Henceforth, we will refer to this sample as Zo17.

\citet{Nandakumar2018} measured chemical abundances for  71 giant stars
in the inner two degrees.
We want to emphasize that the stars in the GC field are stars surrounding the NSC and are not  members of the NSC. 
These chemical abundances have  been obtained by an homogeneous  re-analysis of the high resolution K-band spectra of \citet{ryde:15} and \citet{ryde:16}.  Surface gravities have been obtained in an iterative way (with respect to  $T_{\rm eff}$ and [Fe/H]) by placing them on stellar isochrones. The new  K-band line list (\citealt{Thorsbro2018}) has been also used. Henceforth,  we will refer to this sample as N18.
 Table \ref{survey} shows the summary of the  spectroscopic data set we used in our analysis together with the information about the central coordinates of the  samples, the number of stars, the spectral resolution, the wavelength range as well as the main tracer of the population (e.g. M giants, RC stars).

\begin{table}[!htbp]
\caption{Spectroscopic samples used in this analysis. First column gives the reference, second column the central coordinates in galactic longitudes and latitudes, third column the number of used stars, 4th column the spectral resolution, fifth column the corresponding wavelength range in $\mu m$ and the last column the tracer of the stellar population used.}
\begin{tabular}{cccccc}
Dataset&(l,b)&N&R&$\Delta \lambda$ ($\mu$m)&type\\
\hline
R17&(0, 0)&17&24.000&2.11--2.40&M giants\\
N18&(0, 0)&9&50.000&2.08--2.14&M giants\\
   --  &(0, $\pm 1$)&23   & --    &  --      & --     \\
   -- &(0, $\pm 2$)&  24 & --   & --        & --      \\
   --&(0, 3)& 15 & --  &--             &--      \\
Za19&($\pm 0.2$, $\pm 1$)&13&22.500&1.50--1.70& M giants\\
--  & ($\pm 0.2$, $\pm 2$) &58& --& --& --\\
--&($\pm 0.2$, $\pm 3$) &35& --& --& --\\
Zo17&(0, --1)&432&6500&0.82--0.94 &RC stars\\
--  &(0, --2)&432& --& --  &  --\\
FK17&(0, 0)&706&4000&1.93--2.46& M giants\\
\hline
\end{tabular}
\label{survey}
\end{table}

\subsection{Caveats about chemical abundances from high and low-resolution spectra}

There is a general debate on whether and eventually how chemical abundances from 
low-resolution spectra  ($R\le 5000$) can be compared with those from medium ($R\sim 10,000$) and high-resolution ($R\ge 20,000$). 
In the inner two degrees of the Milky Way that issue is even more critical, given that we are often dealing with metal rich, cool giants suffering from substantial molecular lines in their spectra,  with temperatures well below 4500 K and $\rm log\, g <  2$. Appearance of molecular lines and consequent line blending and blanketing, saturation, inaccurate line lists, uncertainties in stellar parameters, etc.  can severely affect the derived chemical abundances of M giants (see e.g. \citealt{ryde:16}, \citealt{Nandakumar2018}).  In addition to all of these problems, some lines are strongly temperature sensitive due to ionization issues and hyperfine splitting \citep{Thorsbro2018}.

\begin{table}[!ht]
    \caption{Derived temperature, gravity and metallicity for a test giant star in common to different studies.}
    \label{tab:obsstar}
    \centering
    \begin{tabular}{lccc}
         & Do et al. 2015 & FK17 & This work \\\hline
        T$_\text{eff}$ & $3558 \pm 414$ K & $3479 \pm 229$ K & $3200 \pm 200$ K \\
        log$g$ & $3.0 \pm 0.91$ & $0.1 \pm 1.0$ & $0.3 \pm 0.3$ \\
        $[$Fe/H$]$ & $0.80 \pm 0.32$ & $0.3 \pm 0.33$ & $0.2 \pm 0.2$\\
    \end{tabular}
\end{table}
 
In order to provide a first check of the possible mismatch among stellar parameters  obtained form spectra at different resolutions, on 30 July 2017 at Keck II,
we obtained a high resolution ($R\sim23,000$)  K-band spectrum of the star Nr. 44 from FK17,  that is also in common with \citet{Do:2015}. 
The star is at $(RA,DEC) = (266.42050\degr,-29.006775\degr)$ using epoch J2000 and it has $K_s = 10.4$\,mag.
We use the Spectroscopy Made Easy (SME) code for spectral synthesis \citep{sme,sme_code} and  the MARCS model atmospheres \citep{marcs:08} in spherical geometry.

\begin{figure}[!h]
    \centering
    \includegraphics[width=0.49\textwidth]{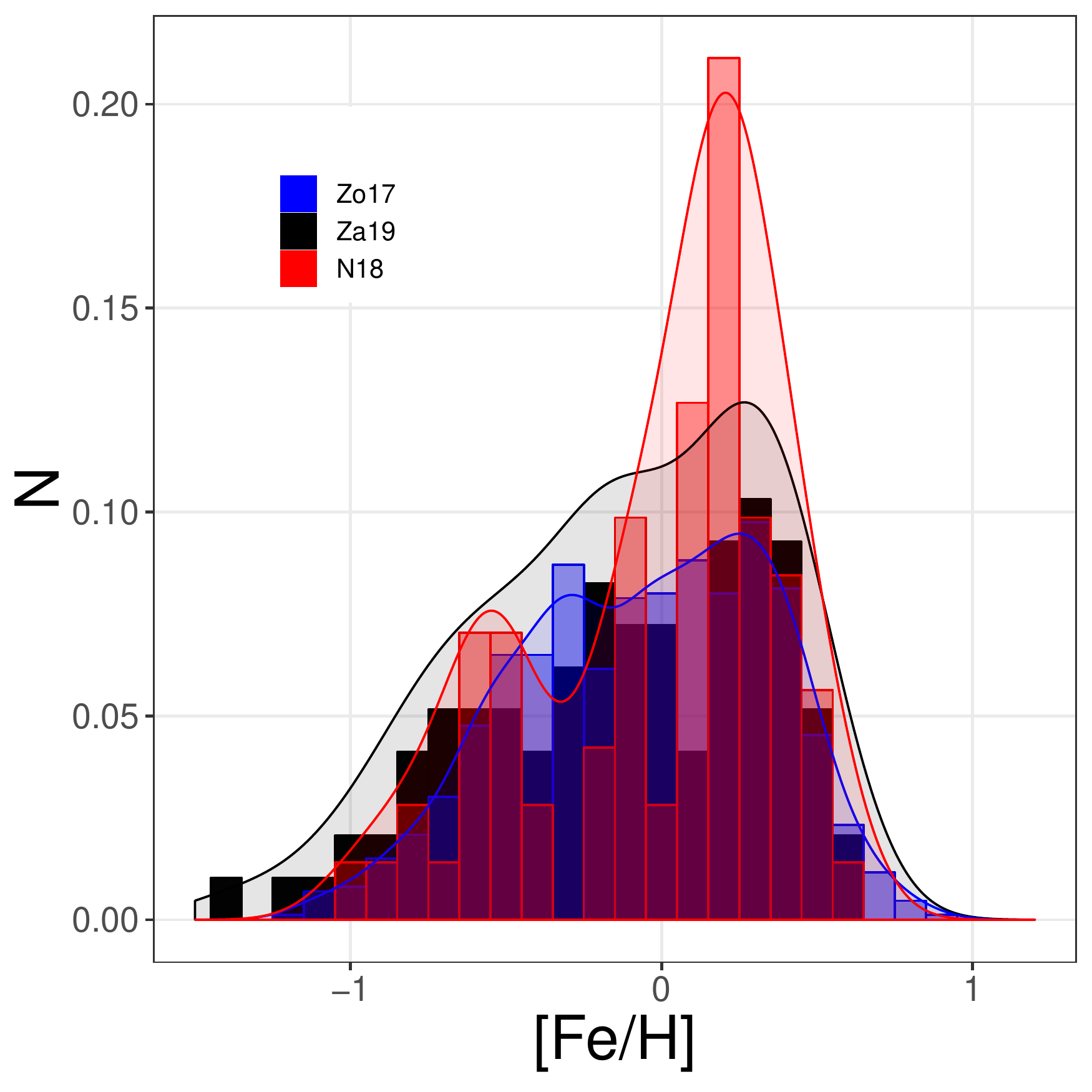}
    \caption{Normalized Metallicity Distribution Function of M giants between the individual samples of Za19 (black), Zo17 (blue) and N18 (red).
A kernel density estimation (KDE) for each of the three samples with a binsize of 0.1\,dex is  overplot as solid line.
}
    \label{hist_FeH_comparison}
\end{figure}

\begin{figure}[!h]
    \centering
    \includegraphics[width=0.49\textwidth]{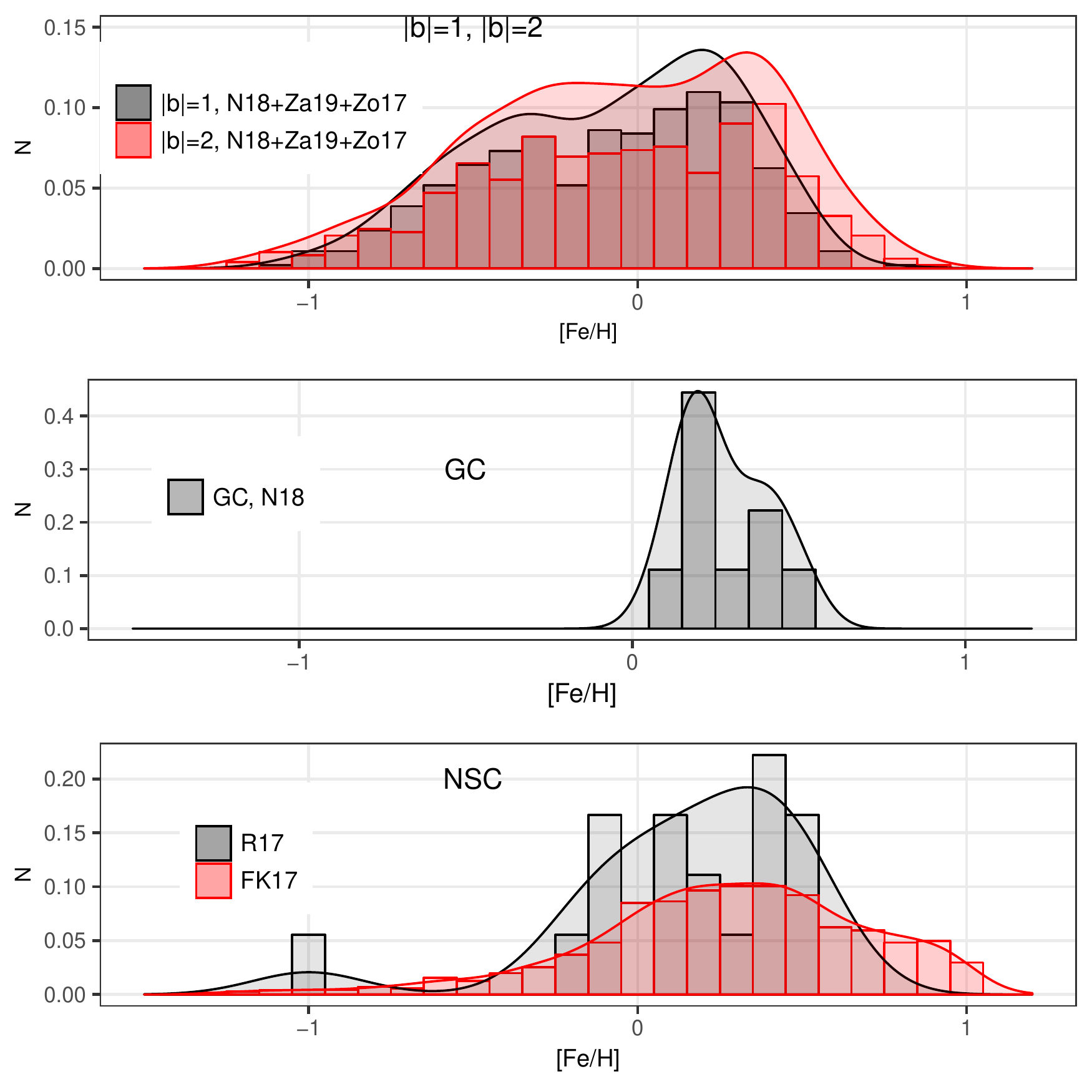}
    \caption{Normalized Metallicity Distribution Function of M giants in the inner two degrees. Upper panel shows the distributions at |b| = 1 deg and |b| = 2 deg from Zo17, N18 and Za19, middle panel the one for the GC field from N18 and the  lower panel for the NSC from FK17 and R17. 
A kernel density estimation (KDE) with a binsize of 0.1\,dex is also overplot as solid line.
}
    \label{hist_FeH_all}
\end{figure}

Table~\ref{tab:obsstar} shows our best-fit stellar parameters and metallicity together with the corresponding values found by \citet{Do:2015} and FK17.
Our values of the stellar parameters and metallicity for  this star are in reasonable agreement with those  of  FK17 within the uncertainties, while both gravity and [Fe/H] by
\citet{Do:2015} seem unlikely too large. For this reason, as well as the previously mentioned issue with unphysical gravities and extreme metallicities,  we choose to exclude the R=5000 sample of \citet{Do:2015}  from our analysis. 

\section{The metallicity distribution function}

By using the state-of-the-art metallicity samples described in Sect.~\ref{samples}, we computed  the MDF in
the inner two degrees of the MW. Fig.~\ref{hist_FeH_comparison}  shows the individual MDFs of Zo17, Za19 and N18, respectively. Their median metallicity is -0.04, -0.06 and 0.04\,dex, respectively with a typical r.m.s of 0.4\,dex, showing that individual MDFs are comparable.  In order to maximize the statistical significance of each distribution, we merged the Zo17, N18 and Za19 samples in the inner bulge fields which is shown in Fig.~\ref{hist_FeH_all}. We refer to this as the ``merged sample''.

Figure~\ref{hist_FeH_all} shows also the normalized MDFs for the inner bulge fields at  $\rm b = \pm 2\degr$ and $b = \pm 1\degr$ (upper panel), for the GC (middle panel) and for the NSC (bottom panel).
The MDF of the inner bulge fields (Zo17, N18, Za19) is very broad, with some evidence of bimodality  as for  bulge fields at higher galactic latitudes (e.g. Baade's window). However the two peaks at super-solar and sub-solar metallicity are less pronounced in the inner bulge merged sample.
The  MDF in the GC is narrow with a single peak at super-solar metallicity and no metal-poor stars. However, we want to stress that the GC sample  consists of only nine stars and a larger sample is clearly needed before ruling out the presence of metal-poor stars. Future large surveys in the IR such as MOONS at the ESO/VLT \citep{ciras16} will definitely increase the sample by at least an order of magnitude.
The MDF in the NSC is comparable to the  GC but broader, with a tail toward sub-solar metallicities. 
In R17 the metal-poor tail is less extended than in FK17 but  the sample size of R17 is too small to  do a more detailed quantitative analysis.   In order to merge the R17 and the FK17 sample, we performed Monte-Carlo simulations where from the FK17 sample we randomly extracted 17 stars (which is the sample size of R17) 10.000 times. The resulting mean and standard deviation of the extracted samples are  within 0.05\,dex in [Fe/H] of the full R17 sample.
In the following we will refer to the GC and the NSC as the GC region.

\begin{figure}[!h]
    \centering
    \includegraphics[width=0.49\textwidth]{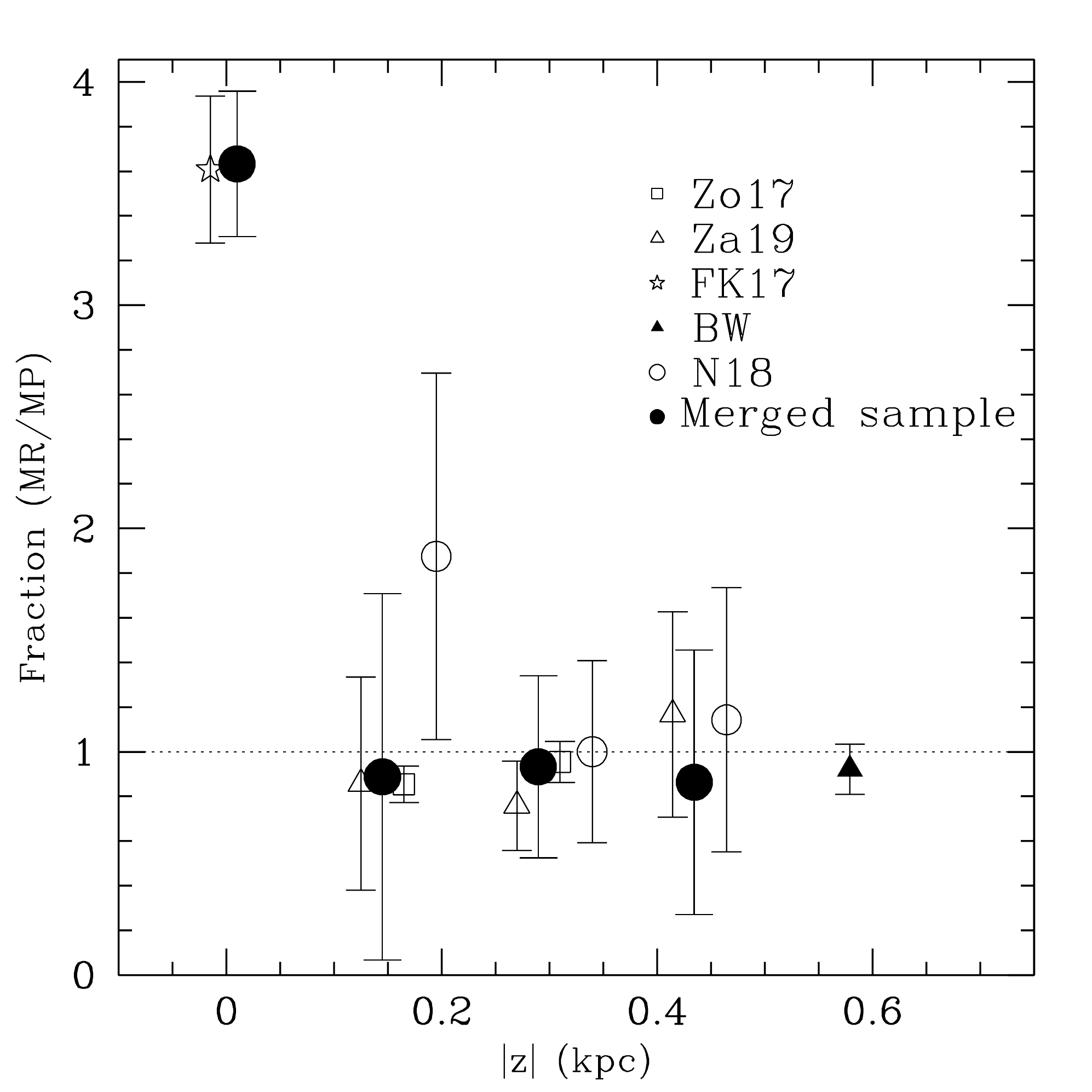}
    \caption{Fraction of MR ([Fe/H] > 0) to MP stars  ([Fe/H] < 0) as a function of the vertical distance for the considered samples of stars. Errorbars have been computed as $\rm {\Delta~f= f\times\sqrt{(1+f)/MR}}$, where f is the number ratio of  MR to MP stars and MR is the number of MR stars}.
    \label{Fraction}
\end{figure}

The most striking difference between the MDF in the inner bulge and in the GC region  is
the larger population of metal-poor stars in the former. 
This difference is significant. Indeed, 
a statistical KS-test gives a $p$-value of 0.02, indicating that these two distributions do not belong to the same population. 
Such a difference would favour a scenario of a in-situ formation of the NSC. 
However, infall of complex stellar systems such as Terzan 5 (\citealt{Origlia2013}, \citealt{Origlia2019}) with major age and metallicity spreads or 
genuine globular clusters \citep{Capuzzo-Dolcetta2008,Tsatsi2017} cannot be fully ruled out as possible contributors. 
In the latter case, chemical footprints of dispersion in light elements \citep{Schiavon2017} should be detected and chemical abundances for  
these metals are urgently needed.

\section{Metallicity gradients}

In order to consistently trace possible vertical and radial gradients in the Galactic bulge, it is important to quantify the relative contribution of the super-solar, metal-rich (hereafter MR) and the sub-solar, metal-poor (hearafter MP) stars. Indeed, when the MDFs are complex as in the Galactic bulge, a straightforward average value is not always appropriate to describe the metal content of the 
sampled stellar population at a given distance from the GC.

We thus divide each sample described in Sect.~\ref{samples} into two components, i.e.  the MR with  $\rm [Fe/H]>0$ and the MP with  $\rm [Fe/H]<0$, following \citet{zoccali2017}. 
Then we compute the number ratio of MR to MP stars and their average metallicities.
We verified that slightly (by $\pm$0.2 dex from [Fe/H]=0.0) different assumptions to separate MR and MP stars do not significantly affect the results.  In this respect, \citet{Nandakumar2017} already demonstrated that the  effect of the selection function plays a minor role in MDF studies.

\subsection{The vertical metallicity gradient}

\begin{figure}[htbp!]
    \centering
    \includegraphics[width=0.49\textwidth]{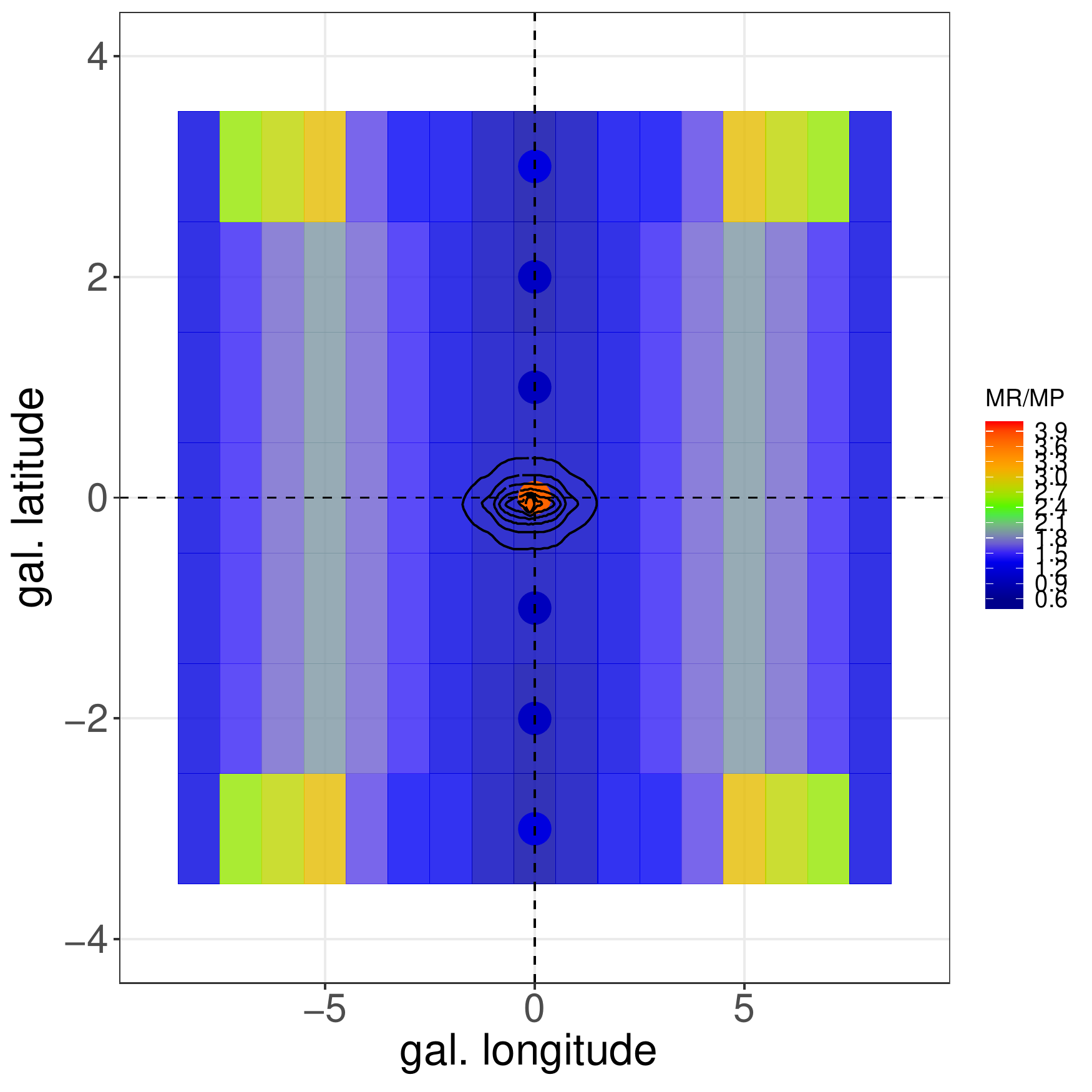}
    \caption{Fraction of  the metal-rich to metal-poor stars (filled circles) superimposed by  the  fraction of MR/MP stars from Zoccali et al. (2018, Tab. 1).  The black contours show the surface brightness map of the best fit model of the  "nuclear bulge" component by \protect \citet{launhardt2002}.}
    \label{frac_zoc}
\end{figure}

By using the same state-of-the-art metallicity samples as for the MDFs, we derived number ratios of MR to MP stars and their average metallicities,  and we investigate their behavior as a function of the distance (z) from the Galactic plane. Here we use for the merged sample at z=0 the dataset of R17, FK17 and the stars at the GC from N18. For the merged sample of |b|=1\degr, |b=2|\degr  and |b|=3\degr,  we use the corresponding datasets of N18, Za18, and Zo17 as described in Sect.~3.
Figure~\ref{Fraction} shows the fraction of  MR to MP stars with varying $z$. For comparison we also 
include the corresponding values at $\rm b = \pm 3$ deg from \citet{Nandakumar2018} and \citet{Zasowski2019} and in  Baade's window (BW) from \citet{schultheis:17}.  
The fraction of MR to MP stars for $\rm |b| \geq 1^{o}$ (corresponding to z=0.14\,kpc)  out to the BW  at z=0.57\,kpc turns out to be rather constant around the value of 1, suggesting an almost equal fraction of MR and MP stars \citep[see also][]{zoccali2017}.
At variance, in the GC region the MR/MP number ratio significantly increases up to a value of about 3.6, indicating a dominance of the MR component.  Figure~\ref{frac_zoc} shows the map of the 
MR/MP number ratio in the innermost bulge region while in Tab.~\ref{ratioMRMP} we list for each field the fraction of MR/MP stars.   For comparison we also show the ratio of MR to MP stars from \citet{Zoccali2018}  based on the GIBS data (see their Tab.~1) but confined to $\rm |b| \leq 3\degr$.
The distribution is rather flat, with values ranging between 0.8 and 1.2 with a spike in the GC region. We superimpose on Fig.~ \ref{frac_zoc} the stellar mass distribution of the nuclear
disc from \citet{launhardt2002} where our metallicity peak
corresponds with the concentration of the mass within the nuclear disc/NSC. Subsequent studies all confirm the extreme concentration of stellar mass in the nuclear disk and the NSC (e.g. \citealt{fritz16})
Interesting enough, the individual samples and the merged ones (i.e. FK17+R17 in the GC region, N18+Za19+Zo17 in the $\rm b = \pm 1$ and 
$\rm b = \pm 2$ fields, N18 + Za19 in the $\rm b = \pm 3$ field)
show similar MR/MP number ratios, despite their different spectral resolution,  line diagnostics and statistical significance.
This suggests that the merged samples, i.e. those with the highest statistical significance, can be safely used to trace possible metallicity variations.
Our finding that different spectral resolution and line diagnostics provide similar results is not surprising, given that  we are simply considering  two broad metallicity categories (i.e. MR and MP).

\begin{table}[!htbp]
\caption{ Ratio of MR to MP stars for each of the spectroscopic samples as well as the merge samples. Errorbars have been computed as $\rm {\Delta~f= f\times\sqrt{(1+f)/MR}}$, where f is the number ratio of  MR to MP stars and MR is the number of MR stars} 
\begin{tabular}{llll}
Dataset&(l,b)& ratio (MR/MP)& $\rm \sigma_{(MR/MP)}$\\
\hline
FK17&(0, 0)&3.61&0.30\\
FK17+R17+N18&(0, 0)&3.63&0.33\\
N18&(0, $\pm 1$)&1.87&0.82\\
N18&(0, $\pm 2$)& 1.00 & 0.41  \\
N18&(0, 3)& 1.14 & 0.59    \\
Zo17&(0, --1)&0.85&0.08 \\
Zo17 &(0, --2)&0.95&0.09 \\
Za19&($\pm 0.2$, $\pm 1$)&0.86&0.48\\
Za19& ($\pm 0.2$, $\pm 2$) &0.76& 0.20 \\
Za19&($\pm 0.2$, $\pm 3$) &1.17& 0.46\\
N18+Zo17+Za19&(0, $\pm 1$)&0.89&0.82\\
N18+Zo17+Za19&(0, $\pm 2$)&0.93&0.41\\
N18+Za19&(0, $\pm 3$)&0.86  &0.59\\
BW&(1, -4)&0.92&0.11\\
\hline
\end{tabular}
\label{ratioMRMP}
\end{table}

\begin{table}[htbp]
    \centering
 \caption{Linear fitting (i.e.  y = a1*x + a0) parameters and 1$\sigma$ dispersions for the vertical metallicity gradient. The overall rms of the residuals is given in the last row. The second column (All) shows the parameters for the full sample, the third one (AllB) for the full sample in the inner bulge without the GC,  fourth and fifth ones for the MR and the MP  sub-samples including the GC, and columns six and seven for the MR and MP populations in the inner bulge without the GC.}
    \begin{tabular}{ccccccc}
Fit &All  & AllB & MRall & MPall & MRB & MPB  \\
            \hline \\
        a0 &  0.099 & -0.042 & 0.345 & -0.281 & 0.254 & -0.367 \\ 
       $\sigma_{a0}$ &0.069  &  0.023 & 0.059 & 0.061 & 0.073 & 0.085 \\
        a1 &  -0.270 & 0.057 & -0.179 & -0.061 & 0.031 &  0.136\\
        $\rm \sigma_{a1}$ & 0.198&  0.058 & 0.165 & 0.174 & 0.182 & 0.212\\
        rms &0.09 & 0.03 & 0.08 & 0.07 & 0.06 & 0.07 \\
        \hline \\ 
    \end{tabular}
    \label{table2}
\end{table}

\begin{figure}[htbp!]
    \centering
    \includegraphics[width=0.49\textwidth]{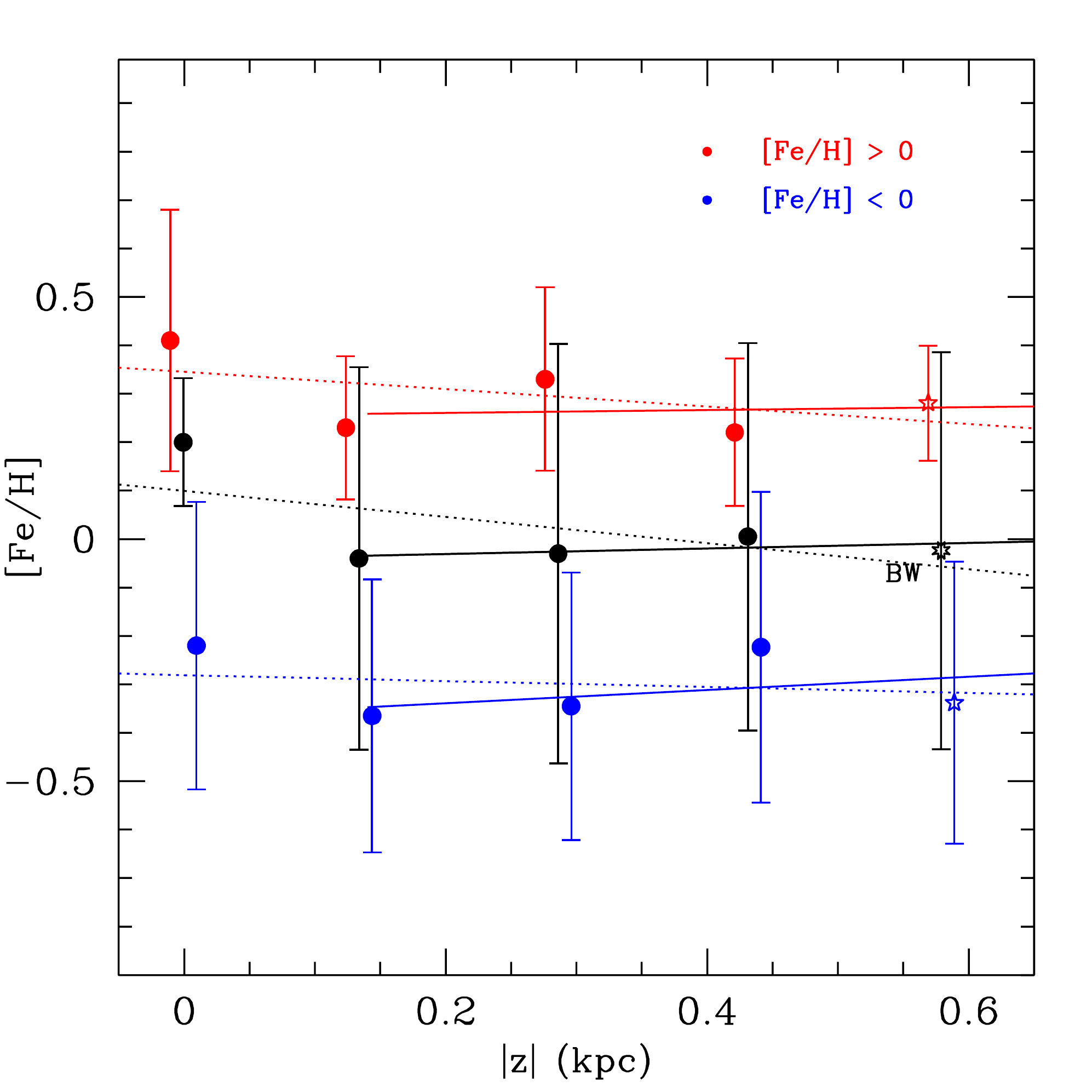}
    \caption{Vertical metallicity gradient  in the inner bulge and in the GC region for MR (red dots), MP (blue dots) and 
all the stars together (black dots) from the merged samples of FK17+R17 in the GC region, N18+Za19+Zo17 in the $\rm b = \pm 1 \degr$ and 
$\rm b = \pm 2 \degr$ fields, N18 +Za19 in the $\rm b = \pm 3 \degr$ field) and of \citet{schultheis:17} in the BW. 
Errorbars refer to 1$\sigma$ dispersions. Dotted lines show the linear fit including the GC region while the solid lines show the fit omitting the GC region.}
    \label{gradient}
\end{figure}

\begin{figure}[h!]
    \centering
    \includegraphics[width=0.49\textwidth]{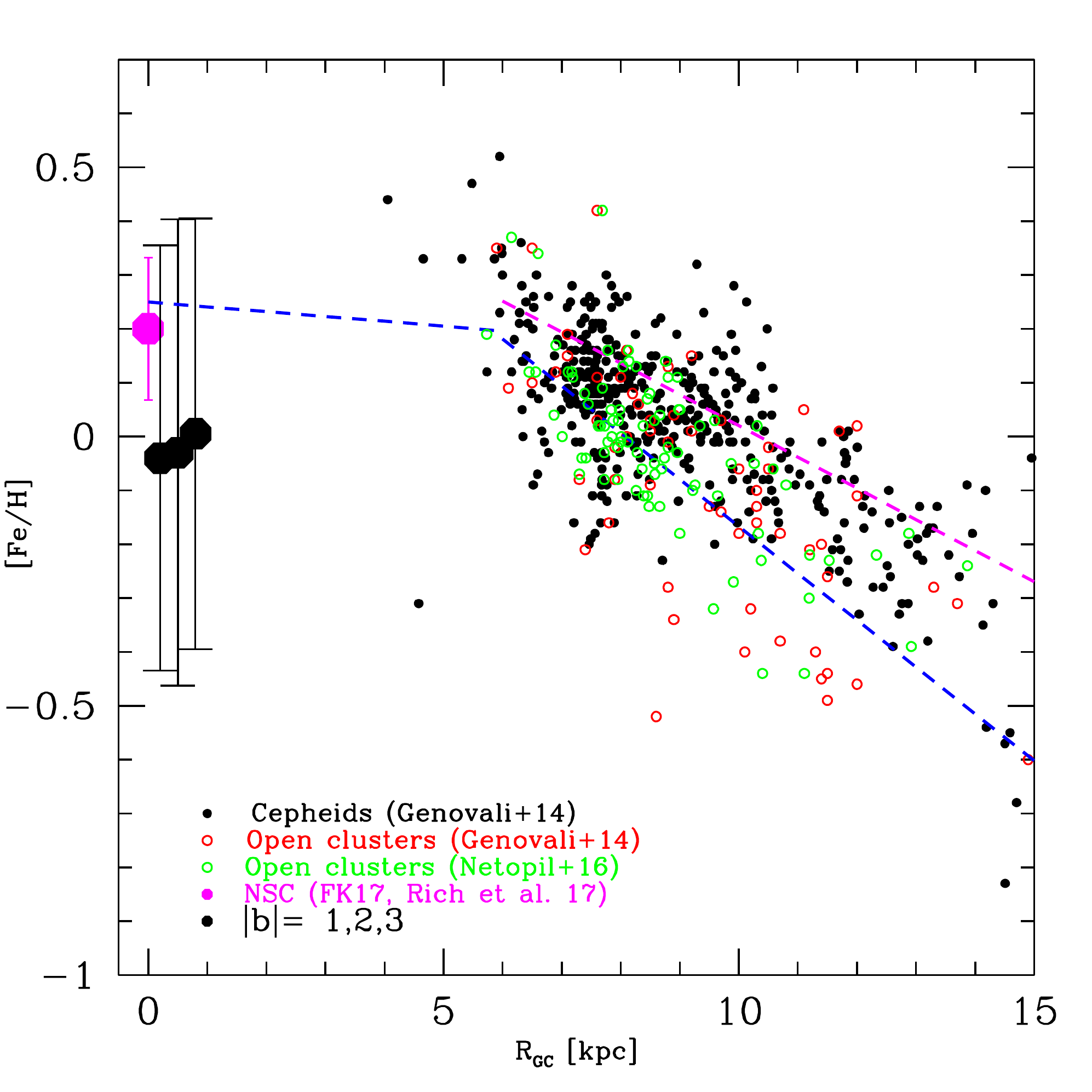}
    \caption{Radial gradient close to the Galactic plane, as traced by 
     Cepheids (black dots from \citealt{Genovali2014}), open clusters (red dots from \citealt{Genovali2014} and green circles from \citealt{netopil16}), big magenta and black dots are the average metallicities for the GC region and |b|=1,2,3 deg inner bulge fields discussed in this work. Errorbars refer to 1$\sigma$ dispersions. The blue dashed line marks the relation by \citet{Hayden2014} for a large sample of APOGEE M giant stars covering a wide  range of Galactocentric distances. The magenta dashed line  marks the relation found by \citet{Anders2017} for red giant stars with well deterimined asteroseismological parameters.}
    \label{radialgradient}
\end{figure}

Figure \ref{gradient} shows the median metallicities  for the individual MR and MP components as well as for the global one, as a function of z.  
We only use the merged samples since they are statistically more significant and we perform a linear square fit to the data points. The coefficients are reported in Table~\ref{table2} .
Very small positive slopes in the inner bulge fields (thus excluding the GC region) are obtained, the highest value  for the MP component, 
indicating that the mild (if any) positive gradient is mostly due to the decrease of the median metallicity of the MP component moving outwards.
At variance, if the  GC region is included, slopes become negative and the mild negative gradients are practically due to the high median metallicity 
of the dominant MR population in the GC region.
Interestingly, all the slopes in Table~\ref{table2} are consistent with flat distributions at $\le 1.5\sigma$ level.

\subsection{The radial metallicity gradient}

The radial metallicity gradient is an important constraint on the chemical evolution for the  Milky Way and external  galaxies. 

Figure~\ref{radialgradient} shows the mean metallicity for open clusters and Cepheids in the range $5 < R_{\rm GC} < 15\,{\rm kpc}$ by 
\citet{Genovali2014} and \citet{netopil16} as a function of the Galactocentric distance and the fitting relations based on observations of 
field giant stars by \citet{Hayden2014} and \citet{Anders2017}. For comparison, mean metallicities for the GC region and the inner bulge fields as 
discussed in this paper are also reported.
There is clear evidence that the extrapolation of the radial gradient inferred at large Galactocentric distances to the center would result in a 
metallicity exceeding the measured central value by $\rm \sim 0.6\,dex$. 
All the measurements in the inner bulge and in the GC region indicate average metallicities between solar and twice solar at most. As already speculated by \citet{Hayden2014}, this flattening might arise due to the mixing of the stars in the presence of the galactic bar.

\section{Alpha-element abundances}

In order to determine individual abundances, high spectral resolution ($\rm R > 20.000$) is necessary. Indeed, very few studies exist in the inner two degrees on detailed chemical abundances of alpha-elements in the bulge low latitude fields. 
\citet{rich:07} did the first detailed chemical abundance analysis and found a
rather homogeneous $\rm \alpha$-enhancement of $\rm \sim 0.3\,dex$ up to solar metallicity for a sample of 17 M giants located at $(l,b)=(0,-1)$. In a similar study \citet{rich:12}  extended this work to two  more fields finding  the same average $\rm \alpha$-element enhancement of $\rm \sim 0.3\,dex$. Also, the study of the metal-poor stars in the NSC ([Fe/H]$\sim -1$) found by \citet{Ryde2016}, shows an $\alpha$-enhancement of $\sim 0.4\,\rm dex$.

\begin{figure*}
    \centering
    \includegraphics[width=0.49\textwidth]{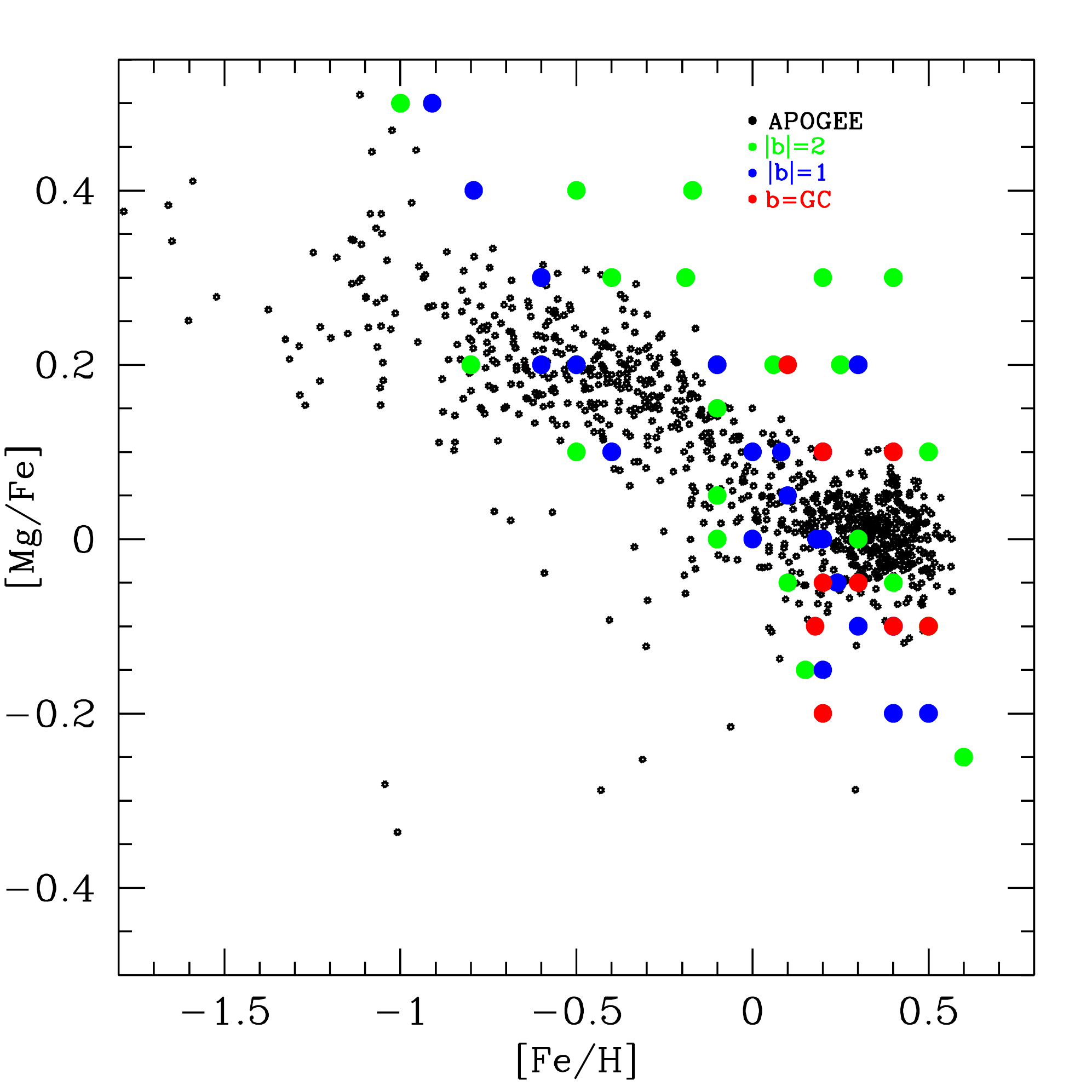} \includegraphics[width=0.49\textwidth]{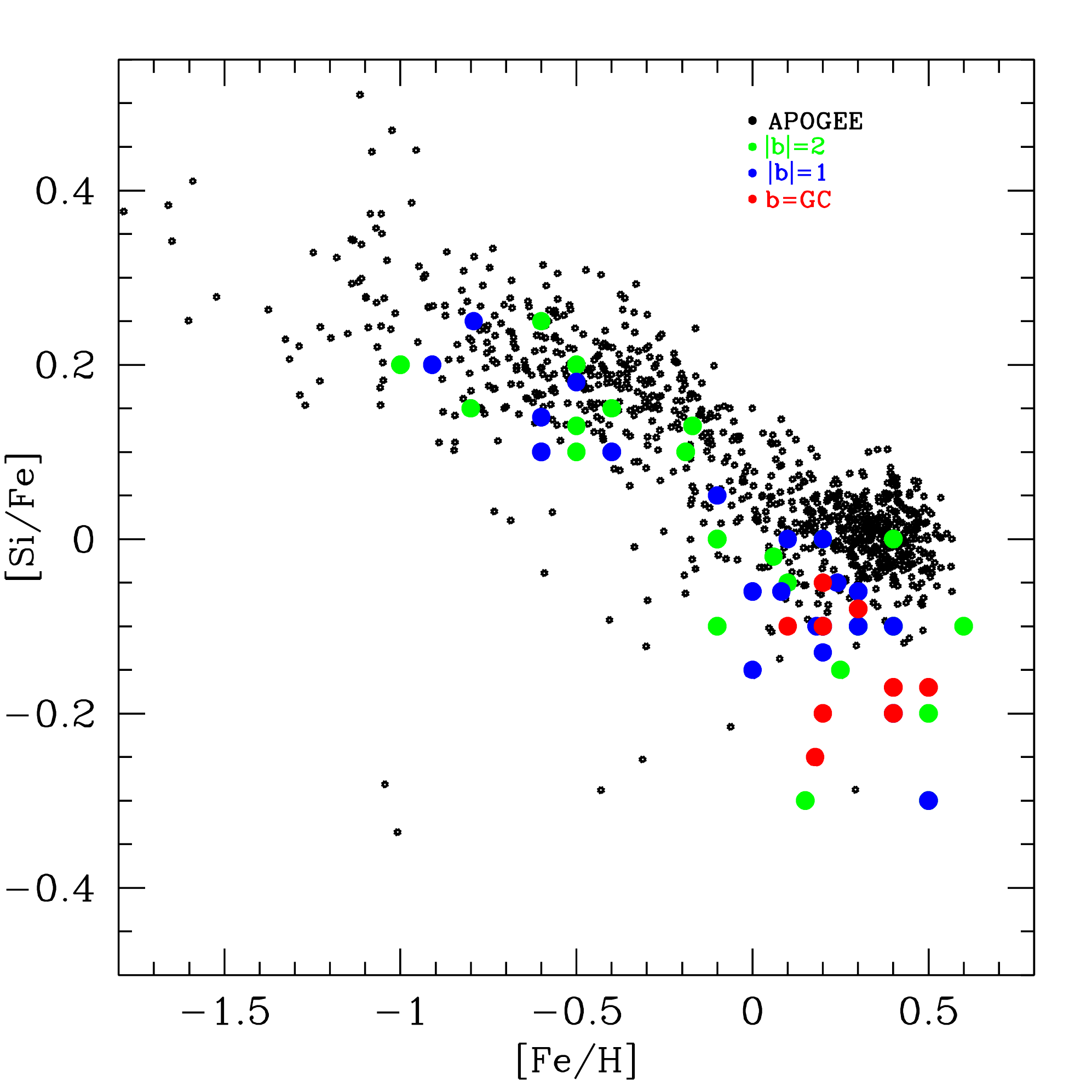}
    \caption{[Mg/Fe] {\it vs} [Fe/H] (left panel) and [Si/Fe] {\it vs} [Fe/H] (right panel) trends in the inner bulge and GC regions.}
    \label{abundances}
\end{figure*}

Fig.~\ref{abundances} shows the [Mg/Fe] and [Si/Fe] abundance ratios as a function of [Fe/H] as obtained by  \citet{Nandakumar2018} in the 
central two degrees and by \citet{Zasowski2019} using 
the latest APOGEE  DR14 release of bulge chemical abundances.  However, we see that the APOGEE
  [Mg/Fe] and [Si/Fe]  abundances level off above $\sim$ solar metallicity, while our $\rm \alpha$-abundances continue decreasing with increasing metallicity. \citet{Matteucci2019} reproduces this flattening by
artificially increasing  the Mg produced by Type Ia SNe by a factor of 10. However, 
we believe that this flattening is unphysical, and results from an artifact in the analysis of metal-rich cool M giants.

In the |b|=1\degr and |b|=2\degr fields we observe a  general decline  of the [Mg/Fe] abundance ratio with increasing metallicity down to roughly solar-scaled [Mg/Fe] at super-solar [Fe/H],  similar to that found by APOGEE, although with larger scatter.
The [Si/Fe] distributions tend to be systematically lower (by 0.1-0.2 dex) than APOGEE, especially  at [Fe/H]>-0.5 dex .
The GC distributions follows those of the |b|=1\degr and |b|=2\degr fields at super-solar [Fe/H]. However, we want to stress that the GC sample is very small and clearly more observations are necessary to do a more quantitative comparison.

\section{Discussion and Conclusions}

Based on high and low-resolution spectroscopic observations of giant stars located in the inner two degrees of the galactic bulge, we find evidence of a dominant 
population of metal rich stars in the GC region, leading to an average metallicity of $\rm +0.2\,dex$.
Moving outward from the GC, in the central 600 pc we find that the ratio of metal rich to metal poor stars is constant, with no evidence of vertical abundance gradients, 
confirming on a more robust statistical basis the previous work of \citet{ramirez:00}, \citet{rich:07}, \citet{rich:12} and \citet{babusiaux:14}.  

The GC region hosts a complex stellar population with a large spread in age and a chemical content that significantly differs from the bulge stellar population in the inner 600 pc. Hence, in order to properly assess the possible presence of vertical or radial gradients in the bulge and inner disc, the GC region should be excluded.

Interestingly, the vertical metallicity gradient of about $\rm -0.3\,dex/kpc$ \citep{Rojas17,gonzalez:13} measured at galactic latitudes $\rm b \geq 4^{o}$ holds also in the inner 600 pc  only if the GC region is included; a proper measurement of the gradient should not consider the GC region.

We can thus confidently state that the metallicity distribution in the inner 600 pc of the Galactic bulge is rather homogeneous. 
Only the innermost $\sim 10$ pc shows an overall higher mean metallicity,  suggesting that the GC region is a very peculiar environment with a large concentration of super-solar  metal-rich stars, perhaps arising from the retention of metal rich gas by the potential well of the NSC.
It is worth noting that also the radial metallicity gradient observed in the disc at Galactocentric distances $>$6 kpc does not extrapolate 
when  moving inwards toward  the low latitude bulge fields and the GC region.  

There have been various mechanisms proposed to explain the origin of the vertical metallicity gradient of the MW bulge. It was originally thought 
that  a dispersion-dominated classical bulge/spheroid was the only possible explanation for the vertical metallicity gradient (e.g. \citealt{zoccali2008}). However, other studies have shown that the redistribution of stars by a bar and a boxy-peanut bulge can also produce a vertical metallicity gradient  (see for example \citealt{Martinez2013}; \citealt{dimatteo:14}). \citet{fragkoudi:17} showed that while such models can indeed reproduce the vertical metallicity gradient, they do not reproduce a number of other trends seen in the MW bulge and inner disc, such as the metal-poor stars in innermost regions and the positive longitudinal gradient. 

Very few chemical evolution models exist in the inner two degrees. \citet{grieco:15} studied the chemical evolution of the central region  and concluded that the stars in the GC formed very rapidly (0.1--0.7\,Gyr) with high star formation efficiency ($\rm 25\,Gyr^{-1}$) and a top-heavy IMF. \citet{fragkoudi:18} used a composite model of a thin and a massive centrally concentrated thick disc that evolves in forming a bar and a boxy/peanut bulge and find a remarkable good agreement with the metallicity distribution functions of APOGEE. They predict a gradient of -0.10\,dex/kpc for the inner 500\,pc.  However,  this model is a simple  closed-box model which does not include star formation and stellar feedback processes nor any gaseous physics.

The presence of RR Lyrae stars in the NSC \citep{Dong2017} favors at least a non-negligible portion of the population being $\sim10$ Gyr or older, a point that is also made by \cite{Nogueras-Lara2018b}.  

Our evidence for a high metallicity peak in the GC might plausibly be explained by a rapid in-situ star formation event in an enriched gas environment, which might have involved coevolution of the SMBH growth and the NSC. 
The formation of multiple stellar populations in the NSC might well be problematic due to the unique environment of SgrA*     (e.g. a strong tidal field from the SMBH) and also at some point in its history would have  included substantial AGN activity.  Further, high energy phenomena are observed at the present day, including magnetic fields and outflows of hot gas \citet{Ponti2019}. It is possible that the conditions that produced the supermassive black hole arose at the same time that vigorous star formation was building the NSC, leaving the fossil remnant of high metallicity that we observe today. However, the strong tidal fields near Sgr A* remain as an unsolved challenge to explaining star formation - the so-called "paradox of youth"
\citep{Ghez2003}.   An attractive idea is to attribute the observed high metallicity spike to enhanced star formation rates in AGN hosts compared to non-AGN that appear to be the case \citep{Santini2012} and appear more pronounced for high luminosity AGNs. On the other hand, \citet{Stanley2015} found no evidence of a correlation  between SFR and AGN activity across all AGN luminosities.

It would be valuable to test this hypothesis by measuring the star formation history and metallicity of other nuclear star clusters, perhaps spanning a range in black hole mass.  It is also important to use the next generation of facilities to constrain in greater detail the star formation of the NSC and surrounding nuclear disk and bulge population.

We have exploited the high luminosities of late M giants to make possible the investigation of the NSC using the current generation of instrumentation. 
Spectrographs coming online in the near future, such as CRIRES+ or MOONS at the VLT, and, later on, HIRES, HARMONI, and MOSAIC at the ELT,
will make possible a deeper exploration of this region that may eventually include the red clump stars.   The NSC offers our best opportunity to explore the coevolution of a stellar population and a supermassive black hole; such studies will be of great interest in the next decade.

\begin{acknowledgements}
We want to thank the referee E. Valenti for her very constructive comments. We want to thank F. Fragkoudi and A. Mastrobueno-Battisti for their very fruitful discussions.
M.S. acknowledges the Programme National de Cosmologie et Galaxies (PNCG) of CNRS/INSU, France, for financial support.  N.R. acknowledge support from the Swedish Research Council, VR (project number 621-2014-5640), Funds from Kungl. Fysiografiska S\"allskapet i Lund. (Stiftelsen Walter Gyllenbergs fond and M\"arta och Erik Holmbergs donation), the project grant “The New Milky” from the Knut and Alice Wallenberg foundation, and support from the Crafoord Foundation, and Stiftelsen Olle Engkvist Byggm\"astare.

Some of the data presented herein were obtained at the W. M. Keck Observatory, which is operated as a scientific partnership among the California Institute of Technology, the University of California and the National Aeronautics and Space Administration. The Observatory was made possible by the generous financial support of the W. M. Keck Foundation. The authors wish to recognize and acknowledge the very significant cultural role and reverence that the summit of Mauna Kea has always had within the indigenous Hawaiian community.  We are most fortunate to have the opportunity to conduct observations from this mountain.
\end{acknowledgements}

\bibliographystyle{aa}
\bibliography{review_revised}

\end{document}